\documentclass[a4paper,fleqn,usenatbib]{mnras}

\usepackage{newtxtext,newtxmath}

\usepackage[T1]{fontenc}
\usepackage{ae,aecompl}
\usepackage{wasysym}

\usepackage{multirow,graphicx}	
\usepackage{amsmath}	
\usepackage{amssymb}	
\usepackage{siunitx}
\usepackage{lineno,hyperref}

\newcommand{\ca}[1]{{\color{black}#1}}

\title[Studying the temperatures of lunar impact flashes.]{Temperatures of lunar impact flashes: mass and size distribution of small impactors hitting the Moon.}

\author[]{
C. Avdellidou$^{1,2}$\thanks{E-mail: chrysa.avdellidou@oca.eu} and J. Vaubaillon$^{3}$
\\
$^{1}$Science Support Office, Directorate of Science, European Space Research and Technology Centre (ESA/ESTEC), 2201 AZ Noordwijk, The Netherlands\\
$^{2}$Universit\'e C\^ote d'Azur, Laboratoire Lagrange, Observatoire de la C\^ote d'Azur, CS 34229-F 06304 Nice Cedex 4, France \\
$^{3}$Institut de Mécanique C\'eleste et de Calcul des Éph\'emerides, Observatoire de Paris/ PSL, 77 Av Denfert Rochereau, 75014, Paris, France\\}

\date{Accepted XXX. Received YYY; in original form ZZZ}

\pubyear{2019}

\begin{document}
\label{firstpage}
\pagerange{\pageref{firstpage}--\pageref{lastpage}}
\maketitle

\begin{abstract}

Lunar impact flashes have been monitored over the last 20 years for determining the mass frequency distribution of near-Earth objects in the cm–dm size range. 
In this work, using telescopic observations in R and I band \ca{from the NELIOTA database}, impact flash temperatures are derived. They are found to range between approximately 1,300 and 5,800 K. In addition, it is also found that temperature values appear to have a distribution significantly broader than a Gaussian function, therefore making it difficult to estimate the impact flash luminous energy by assigning an average temperature. By measuring the flash temperatures and assuming a black body emission, here we derive the energy of the impacts. We also study the potential link of each event to individual meteoroid streams, which allows us to assign an impact velocity and therefore constrain the projectile mass. Impactor masses are found to range between a few to hundreds of grams, while their sizes are just of few centimetres following a size frequency distribution similar to other studies.

\end{abstract}

\begin{keywords}
Moon--meteorites, meteors, meteoroids
\end{keywords}

\section{Introduction}
\label{introduction}

Almost two decades ago the lunar surface started to be monitored with small-aperture telescopes (e.g. 40~cm) for the detection of the light of flashes produced by the impacts of near-Earth objects (NEOs) \citep{ortiz1999,ortiz2000,ortiz2002,ortiz2006, ortiz2015, bouley2012, suggs2008,suggs2014, madiedo2014, madiedo2015, larbi2015}. 
The idea to monitor the lunar surface using photomultipliers was already introduced since the beginning of 90's by \cite{melosh1993}, 
when it was modelled that impact flash events should be detectable on the lunar surface using modest telescopes equipped with photometers, but the impactor size had to be in the order of 1~m and those events are very rare. However, it was not until 2000 when the first impact flash was recorded \citep{ortiz2000}, using ccd cameras instead of photometers. 
The initial purpose of these studies was to measure the flux of impactors on the Moon and extrapolate it to the Earth. Lunar flash monitoring offers the advantage that Moon's surface provides an extended area for detections, whereas the Earth-based sky monitoring systems that search for bolides do not have such a large detection area. Additionally, the Moon represents an ideal impact target as it lacks atmosphere, therefore an impact occurs without the selection effect of a filtering atmospheric medium.

In general, when impacts occur the kinetic energy of the projectile is partitioned: a fraction of said energy is consumed for the excavation of the impact crater and compaction of the target medium, another fraction is used to heat the materials, while the remaining fraction is converted into the kinetic energy of the ejecta. In the case when the impact is energetic enough, the materials of impactor and target are vaporised and plasma can be generated \citep{melosh1993, burchell1996A, burchell1996B, ernst2005}. That depends on the impact speed and the mass of the impacting body in combination with the type of the materials. In order to constrain the energy partition problem, laboratory hypervelocity experiments have been conducted \citep{eichhorn1975,eichhorn1976}. 

Understanding the energy partition problem is essential to convert the measured brightness of a flash to the impacting energy and possibly to the mass of the projectile. Typically, a black body radiator with luminous energy $E_{lum}$ is assumed, which is a fraction $\eta$ (luminous efficiency) of the kinetic energy of the impactor, $KE$,
\begin{equation}
E_{lum}=\eta KE 
\label{eq1}
\end{equation}

The currently largest survey of lunar impact flashes \citep{suggs2014} resulted in the detection of 126 flashes from which they determined a size frequency distribution (SFD) with diameters ranging from 1 to 14~cm, using their own estimation for a velocity-depended value of $\eta$. 
Some previous studies \citep[e.g.][]{suggs2014,ortiz2015}, in order to perform the colour correction to their flash data, assumed an average flash temperature of 2,800~K adopted from the study of \cite{nemtchinov1998} where they modelled the temperature distribution of the vapour generated during an impact of a 10~cm projectile with speed 25~km~s$^{-1}$ and impact angle $30^{\circ}$. 

The advantage of having temperature measurements at several stages of the long-lasting flashes (the cooling of the flash) can give insight to the thermal evolution of an expanding cloud of droplets. This may reveal information for the size of the melted droplets feeding the studies of lunar regolith where melt-droplet spheroids have been found \citep{warren2008,bouley2012}.

A step forward to the lunar observations for impact events was achieved by the ESA-funded NELIOTA project, which is under the responsibility of the National Observatory of Athens. It is the first telescopic survey that is equipped with two cameras, observing simultaneously in two different wavelength bands \citep[R and I, see Section~\ref{instrumentation} and][]{xilouris2018}. This enables the measurement of the temperature of the observed flashes. Using the first flash detections of the NELIOTA project, \cite{bonanos2018} demonstrated a way to derive temperatures of lunar flashes from astronomical observations for the first time. However, at that time the number of available flashes was very limited and the method of temperature calculation was approximated. \ca{In addition to this work, \cite{madiedo2018} provided flash temperature measurement for a flash recorded in 2015 also in two different wavelengths.}

In this work we used the publicly available data from the NELIOTA website and from \cite{xilouris2018}. In the light of the flash magnitude calibration \citep{xilouris2018}, we revised and modified the temperature and mass calculation method compared to the one presented in \cite{bonanos2018}. Our updated temperature and mass derivation method presented here is rigorous. In addition, this larger dataset is used to derive preliminary statistical properties of the flashes and the impacting population. Correlation between flashes and known meteor streams is searched in order to improve the estimation of the impact velocity. 

The instrumentation is briefly presented in Section~\ref{instrumentation} \cite[for further details see][]{xilouris2018}, while in Section~\ref{data} the dataset and its characteristics (e.g. magnitudes, durations, locations) is described. In Section~\ref{results}, together with an attempt to link the impacting bodies to meteor stream populations, are presented the methods to measure the temperatures and the way these were used to calculate masses and sizes of the impacting population. Finally in Section~\ref{discussion} are discussed the findings and implications of this project.

\section{Instrumentation and Observation mode of NELIOTA}
\label{instrumentation}

The data that are used here were obtained with the 1.2~m Kryoneri telescope that was refurbished and is used for the NELIOTA project. The facility is located in Peloponnese, Greece at an altitude of 900~m. It is equipped with two identical Andor Zyla sCMOS cameras and a dichroic beam-splitter, which separates the incoming light at 730~nm and sends each separate beam to a different camera. Between the beam splitter and each camera there is an R$_{c}$ and I$_{c}$ Cousin filter with maximum transmitted wavelengths at 641~nm and 798~nm respectively. The final ranges of the transmitted wavelengths, when the total setup is considered according to the manufacturer's quality tests, are 550--800~nm for the $R-band$ and and 700--950~nm for the $I-band$. In principle, observations should take place every month on the waning and waxing side of the lunar surface. Each night and every 15 minutes of lunar observations photometric standard stars are observed at approximately the same airmass with the Moon. These measurements are used to calibrate the instrumental magnitudes of the flashes \citep{bonanos2018}, which are subsequently published online. Both cameras are synchronised and record with 30 frames-per-second (fps) with the integration time to be $t_{int}=0.023$~s and the readout $t_{r/o}=0.010$~s. The interval between two frames is thus 0.033~s. 

\section{Data}
\label{data}

Detections of lunar impact flashes are reported since February 1$^{st}$ 2017, totalling in $\sim$86~h of observations of both waning and waxing areas of the Moon. After more than 22 months of observations (December 11$^{th}$ 2018) 81 events have been reported of which the 55 are detected in both camera systems and thus are marked as validated. Events detected only in one camera system are reported as candidates. Therefore, the total rate of the validated events is calculated to be 1 flash every $\sim$1.6 observing hours.

Publicly available data of the validated events are downloaded from \emph{neliota.astro.noa.gr} and these are: data and time of each event, their lunar longitude and latitude coordinates, the flash calibrated peak magnitudes in both $R-$ and $I-band$ (in ranges $6.67<R_{mag}<11.8$ and $6.07<I_{mag}<10.4$). However, by comparing the published magnitudes in \cite{xilouris2018} and the ones on the NELIOTA web database\footnote{neliota.astro.noa.gr}, we can see differences, which could affect the temperature estimation and/or its uncertainty. These differences exist because the online magnitudes are approximative and their errors rounded (Liakos and Bonanos personal communication). For this reason for the first 31 validated events the magnitudes and their errors are taken from \cite{xilouris2018}, while for the remaining events the online magnitudes are used. Flash magnitudes are reported in Table~\ref{table1}. In addition, in the same online data archive, the actual datacubes of the events are provided. Each datacube stores the total frames in which a flash appears, together with a few frames after the end of the event.

\section{Methods and results}
\label{results}

\subsection{Determination of time resolved flash photometry}
\label{photometry}

For each flash all frames were extracted from the provided datacube. When a flash appears in more than one frame, instrumental fluxes were derived by Source Extractor algorithms \citep{1996A&AS..117..393B}, applying common parameters in the analysis. In particular, photometry was performed using the FLUX\_AUTO mode and selecting as minimum detected area (DETECT\_MINAREA) the 3 pixels. Additionally, the flux detection threshold (DETECT\_THRESH) was applied to be $1.5\sigma$ above the background RMS and equal to the analysis threshold (ANALYSIS\_THRESH).

The peak magnitudes provided in \cite{xilouris2018} and on the website are already calibrated using standard stars which were observed at similar airmass with the Moon close to the time of the event \citep{bonanos2018} and therefore were used to obtain the calibrated magnitudes of the rest of the frames. Instrumental fluxes of the first frames (both in R and I) and their calibrated magnitudes are used in order to derive the zero-point and thus calibrate the entire sequence of the time resolved flash photometry. For the multi-frame flashes with ID~2 and 13 the magnitudes of all the frames are already presented in \cite{bonanos2018}.

For each flash that the NELIOTA detection algorithm provides, both frames, before and after the event, are inspected manually. If the flash is detected above the aforementioned threshold in $N$ frames the total duration time is estimated as $t=N\times t_{int}+(N)\times t_{r/o}$. Of course the real duration of a flash can be shorter than the number provided by the formula above. But because of the sampling rate used by NELIOTA, any duration shorter than $t_{int}+t_{r}$ = 0.033~s can not be measured. 

Almost 40\% of the flashes last for only a single frame. The validated single-frame events have signal in both $I-$ and $R-band$, while the multi-frame ones have signal in at least two subsequent frames. Both validated and candidate events have signal in the camera system that operates in $I-band$, but the majority of non-validated events lack of detection in $R-band$. In the case of multi-frame events, when the flash is detected in two or more frames in the $I-band$, the detection in $R-band$ is weaker, resulting always in less frames. It is noticed that in some cases the estimated duration, resulting from the photometry in this work, is not in agreement with the duration that is given in the NELIOTA website. In particular in the events with flash ID~17,18, 22, 30 here is detected one frame less. 

\begin{table*}
\centering
\caption{Peak magnitudes, durations and locations of the first 55 flashes as of 11-12-2018, together with the solar longitude at the time of the event. Photometric data and coordinates are taken from NELIOTA webpage and \protect\cite{xilouris2018}.}
\label{table1}
\begin{tabular}{r|crrrrrr}
\hline
\hline
Flash 	& 	Date\_Time 	& Solar Long.	& 	$R \pm \sigma_R$ 	& 	$I \pm \sigma_I$ 	&  	Duration 	& 	Lunar Long. 	& 	Lunar Lat. \\
ID		&	(UT)			&	(deg)	 	&	(mag) 			& 	(mag) 			& 	(s)  		& 	(deg) 		& 	(deg)\\
\hline
1. 	& 	2017-02-01\_17:13:57 	& 312.81    &  10.15 $\pm$ 0.14 	&  	9.05 $\pm$ 0.07	& 	0.033 	&	-29.2		&	-1.5\\
2. 	& 	2017-03-01\_17:08:46 	& 341.07    &	6.67 $\pm$ 0.09  	&  	6.07 $\pm$ 0.07 	& 	0.132 	&	  -9.7		&	-10.3\\
3. 	& 	2017-03-01\_17:13:31 	& 341.08    &	9.15 $\pm$ 0.14 	&  	8.23 $\pm$ 0.10 	& 	0.033 	&	  29.9	&	4.5\\
4. 	&	2017-03-04\_20:51:31 	& 344.23   &	9.50 $\pm$ 0.17 	&  	8.79 $\pm$ 0.10 	& 	0.033  	&	-58.9		&	-12.7\\
5. 	& 	2017-04-01\_19:45:51 	& 12.03	&	10.18 $\pm$ 0.16 	&  	8.61 $\pm$ 0.08 	& 	0.033  	&	-58.8		&	11.6\\
6. 	& 	2017-05-01\_20:30:58 	& 41.41	&	10.19 $\pm$ 0.20 	&  	8.84 $\pm$ 0.09	& 	0.066	&	-43.2		&	4.7 \\
7. 	& 	2017-06-27\_18:58:26 	& 96.07     &	11.07 $\pm$ 0.34 	&	9.27 $\pm$ 0.11 	& 	0.066	&	-22.5		&	26.8 \\
8. 	& 	2017-06-28\_18:45:25 	& 97.01	&	10.56 $\pm$ 0.39 	&	9.48 $\pm$ 0.14 	& 	0.066	&	   0.0		&	5.62\\
9.	& 	2017-07-19\_02:03:36 	& 116.37    &	11.23 $\pm$ 0.41 	&	9.33 $\pm$ 0.10 	& 	0.066  	&	35.0		&	7.8\\
10. 	& 	2017-07-28\_18:21:44 	& 125.62    &	11.24 $\pm$ 0.35 	& 	9.29 $\pm$ 0.09 	& 	0.066	&	-40.0		&	-3.2 \\ 
11. 	& 	2017-07-28\_18:42:58 	& 125.63    &	10.72 $\pm$ 0.26 	& 	9.63 $\pm$ 0.12 	& 	0.033	&	-30.6		&	28.5 \\ 
12. 	& 	2017-07-28\_18:51:41 	& 125.64    &	10.84 $\pm$ 0.26 	& 	9.81 $\pm$ 0.12 	& 	0.033 	&	-50.7		&	20.6 \\  
13. 	&	2017-07-28\_19:17:18 	& 125.66    &      8.27 $\pm$ 0.11 	& 	6.32 $\pm$ 0.09 	& 	0.165  	&	-18.7		&	18.1\\
14. 	&	2017-08-16\_01:05:46 	& 143.13    &	10.15 $\pm$ 0.20 	& 	9.54 $\pm$ 0.12	& 	0.066	&	47.5		&	32.0\\
15.	&	2017-08-16\_02:15:58 	& 143.18    &	10.69 $\pm$ 0.29 	& 	9.11 $\pm$ 0.07	& 	0.066	& 	68.1		&	6.7\\
16.	&	2017-08-16\_02:41:15 	& 143.20    &	10.81 $\pm$ 0.31 	& 	9.08 $\pm$ 0.07	& 	0.066	& 	34.6		&	-15.6\\
17. 	&	2017-08-18\_02:02:21 	& 145.09    &	10.92 $\pm$ 0.21 	& 	9.20 $\pm$ 0.10	& 	0.099	& 	57.8		&	-25.9\\
18. 	&	2017-08-18\_02:03:08 	& 145.09    &	10.19 $\pm$ 0.16 	& 	8.83 $\pm$ 0.10	& 	0.099	& 	76.8		&	13.5\\
19. & 	2017-09-14\_03:17:49 	& 171.27    &	9.17 $\pm$ 0.10 	& 	8.07 $\pm$ 0.04	& 	0.099	& 	70.0		&	-1.1\\
20. & 	2017-09-16\_02:26:24 	& 173.18    &	8.52 $\pm$ 0.10 	& 	7.04 $\pm$ 0.07	& 	0.231 	& 	52.5		&	24.7\\
21. & 	2017-10-13\_01:54:21 	& 199.68    &	9.28 $\pm$ 0.13 	& 	8.37 $\pm$ 0.06	& 	0.132 	& 	65.2		&	-17.3\\
22. & 	2017-10-13\_02:33:43 	& 199.71	&	10.31 $\pm$ 0.25 	& 	9.89 $\pm$ 0.12	& 	0.066 	& 	66.5		&	-12.5\\
23. & 	2017-10-16\_02:46:45 	& 202.69	&	10.72 $\pm$ 0.20 	&	 9.46 $\pm$ 0.11	& 	0.099 	& 	72.5		&	-25.4\\
24. & 	2017-10-26\_17:59:42 	& 213.27	&	10.03 $\pm$ 0.27 	& 	9.42 $\pm$ 0.15	& 	0.033 	&	-33.8		&	-27.9\\
25. & 	2017-11-14\_03:34:15 	& 231.71	&	10.31 $\pm$ 0.19 	& 	9.31 $\pm$ 0.09	&	 0.066 	& 	64.4		&	-29.5\\
26. & 	2017-11-23\_16:17:33 	& 241.33	&	10.45 $\pm$ 0.25 	& 	10.06 $\pm$ 0.14	& 	0.066 	& 	-30.5		&	-35.0\\
27. & 	2017-12-12\_02:48:08 	& 260.02	&	10.50 $\pm$ 0.25 	& 	8.89 $\pm$ 0.10	& 	0.066 	& 	74.0		&	9.0\\
28. & 	2017-12-12\_04:30:00 	& 260.09	&	10.58 $\pm$ 0.29 	& 	9.84 $\pm$ 0.12	& 	0.033 	& 	51.2		&	5.4\\
29. & 	2017-12-13\_04:26:57 	& 261.10	&	10.56 $\pm$ 0.24 	& 	9.95 $\pm$ 0.07	& 	0.033 	& 	50.0		&	13.0\\
30. & 	2017-12-14\_04:35:09 	& 262.13	&	7.94 $\pm$ 0.10 	& 	6.76 $\pm$ 0.11 	& 	0.099 	& 	73.4		&	-36.9\\
31. & 	2018-01-12\_03:54:03 	& 291.64	&	10.01 $\pm$ 0.17 	& 	9.31 $\pm$ 0.10 	& 	0.066 	& 	79.2		&	-40.7\\
32. &		2018-03-23\_17:24:19	& 2.78	&	9.90 $\pm$ 0.30	&	8.6 $\pm$ 0.10		&	0.033	&	-52.0		&	-1.4\\
33.	&	2018-04-10\_03:36:57	& 19.97 	&	8.80 $\pm$ 0.10	&	8.1 $\pm$ 0.10		&	0.033 	&	74.5		&	21.7\\
34.	&	2018-06-09\_02:29:18	& 77.98 	&	9.90 $\pm$ 0.20	&     9.00 $\pm$ 0.10		&	0.033	&	24.6	 	&	4.3\\	
35.	&	2018-06-19\_19:12:09	& 88.20 	&	9.90 $\pm$ 0.20 	&     9.00 $\pm$ 0.10		&	0.033	&	3.6		&	-59.0\\	
36.	&	2018-06-19\_20:00:48	& 88.23 	&      9.90 $\pm$ 0.30	&	9.30 $\pm$ 0.10	&	0.033	&	17.4		&	-58.2\\	
37.	&	2018-06-19\_20:04:09	& 88.23 	&	10.3 $\pm$ 0.60	&	8.60 $\pm$ 0.10	&	0.033	&	2.5		&	-20.0\\	
38.	&	2018-07-09\_01:44:19	& 106.58	&	11.2 $\pm$ 0.30	&	8.60 $\pm$ 0.10	&	0.033	&	46.0		&	24.9\\	
39.	&	2018-08-06\_01:57:43	& 133.33	&	9.70 $\pm$ 0.20	&	8.1 $\pm$ 0.05		&	0.099	&	10.6		&   -22.1\\	
40.	&	2018-08-06\_02:38:14	& 133.36	&	9.20 $\pm$ 0.10	&	7.7 $\pm$ 0.05		&	0.099	&	67.2		&   28.8\\	
41.	&	2018-08-07\_01:33:54	& 134.27	&	10.80 $\pm$ 0.30	&	9.3 $\pm$ 0.10		&	0.033	&	52.1		&   1.8\\	
42.	&	2018-08-07\_01:35:45	& 134.27	&	8.80 $\pm$ 0.10	&	7.7 $\pm$ 0.05		&	0.132	&	70.0		&   3.1\\	
43.	&	2018-08-07\_02:33:18	& 134.31	&	10.10 $\pm$ 0.20	&	9.5 $\pm$ 0.10		&	0.033	&	60.2		&   26.7\\	
44.	&	2018-08-07\_03:10:33	& 134.34	&	10.40 $\pm$ 0.30	&	9.8 $\pm$ 0.10		&	0.033	&	30.6		&   10.3\\	
45.	&	2018-08-08\_02:19:55	& 135.26	&	11.10 $\pm$ 0.30	&	9.9 $\pm$ 0.10		&	0.033	&	34.9		&   21.9\\	
46.	&	2018-08-08\_02:28:23	& 135.27	&	11.10 $\pm$ 0.20	&	10.4 $\pm$ 0.10	&	0.033	&	76.4		&   28.0\\	
47.	&	2018-08-08\_02:29:44	& 135.27	&	8.40 $\pm$ 0.05	&	7.3 $\pm$ 0.10	    	&	0.165	&	60.2		&   26.6\\	
48.	&	2018-08-08\_02:52:25	& 135.28	&	11.10 $\pm$ 0.30	&	9.7 $\pm$ 0.10	    	&	0.033	&	17.4		&   -58.2\\	
49.	&	2018-08-15\_18:08:16	& 142.61	&	11.80 $\pm$ 0.40	&	9.6 $\pm$ 0.10	    	&	0.033	&	-62.4		&   11.7\\	
50.	&	2018-09-04\_01:33:52	& 161.23	&	9.90 $\pm$ 0.30	&	9.2 $\pm$ 0.10	   	 &	0.033	&	 29.2		&   -24.7\\
51.	&	2018-09-05\_01:51:37	& 162.22	&	7.80 $\pm$ 0.10	&	6.6 $\pm$ 0.05	   	 &	0.396	&	52.1		&   9.5\\
52.	&	2018-09-05\_02:47:54	& 162.25	&	10.60 $\pm$ 0.40	&	9.1 $\pm$ 0.10	   	 &	0.066	&	15.2		&   -15.5\\
53.   &   	2018-09-06\_02:00:33	& 163.19    &   11.00 $\pm$ 0.30	&   10.3 $\pm$ 0.10     	&   	0.033   	&	72.5		&   -18.6\\
54.   &   	2018-09-06\_03:10:04	& 163.24    &   11.20 $\pm$ 0.30	&   9.90 $\pm$ 0.10     	&   	0.066   	&	60.8		&   0.0\\
55.   &   	2018-10-15\_18:17:49	& 202.09    &   9.60 $\pm$ 0.20		&   8.80 $\pm$ 0.10     	&   	0.066   	&     -53.3		&	5.5\\
\hline
\hline
\end{tabular}
\end{table*}

\subsection{Determination of possible shower for each impact}
\label{origin}

While flash photometry allows us to derive the left hand side of Eq.~\ref{eq1}, it is important to have a good estimate of the impact velocity in order to derive the mass of the impactor. The link of a flash with a specific meteor shower or the sporadic population has a fundamental role for said problem, because each meteor shower impacts the Moon with a specific and known speed. This allows to having a good choice for the value of $v$ in the righthand side of Eq.~\ref{eq1}. Moreover, the assignment to a meteor shower (or not) could provide information about the density of its bodies. Density information is required to convert masses to sizes and therefore estimate a reliable SFD of the impactor population. The goal of this section is to provide constrains to identify the meteor shower associated to each impact. Using the available dates and times of each flash detection the solar longitude is determined (Table~\ref{table1}). The idea is to check from geometrical considerations, whether a given shower can be responsible for a given impact. A few previous studies have also provided ways to make these links \citep[e.g.][]{suggs2014, madiedo2015}. 

The IAU confirmed meteor shower data\footnote{http://pallas.astro.amu.edu.pl/~jopek/MDC2007, accessed on Nov. 2018.} are considered here. For each impact and each possible parent showers, the different steps to determine whether the shower is a plausible candidate are the following:
\begin{itemize}
    \item We check whether or not there was a shower outburst at the time of the impact.
    \item In turn, we compute the difference in solar longitude ($\lambda_{\odot}$) between the time of the impact and the time of maximum of the shower, or the time of the shower outburst. We check if the difference in solar longitude between the time of the impact and the maximum of the shower (or shower outburst) is smaller than a chosen $\Delta \lambda^{c}_{\odot}$. If not, the shower is dismissed as a possible parent shower.
    \item We compute right ascension and declination $(\alpha,\delta)_{shw}$, which is the position of the shower radiant in inertial coordinates (ICRF at J2000), at the time of the impact, considering the radiant drift.
    \item The location of each impact (longitude, latitude, altitude) is converted into a 3D-position Moon-centred coordinates vector $(x,y,z)$. This $(x,y,z)$ is converted into Moon-centered inertial (ICRF at J2000) coordinates vector $(x,y,z)_{MCI}$, at the time of the impact, and then into spherical coordinates $(r,\alpha,\delta)_{imp}$.
    \item Then, the angular distance $\Delta \gamma$ between $(\alpha,\delta)_{shw}$ and $(\alpha,\delta)_{imp}$ is computed. If $\Delta \gamma$ is larger than a chosen $\Delta \gamma^{c}$ threshold, the shower is dismissed as a possible parent shower.
\end{itemize}
If the shower has made it to this last step, it is considered as a potential parent shower of the impact. This method has the advantage to quickly provide shower candidates for a given impact. At the last step, given the difference in solar longitude $\Delta \lambda_{\odot}$ and the activity strength of the shower ZHR, we derive the most plausible shower for several impacts, listed in table \ref{table2}. For that we used the expression (Eq.~1) as described in \cite{suggs2014}.

It is worth mentioning the limitations of such a method. Firstly, the shower radiants are those visible from the Earth. The gravitation of the Earth and the Moon being different, the effective radiant as seen from the Moon might be slightly different. This effect is especially true for slow meteors, but is ignored here.  The $\Delta \lambda^{c}_{\astrosun}$ and $\Delta \gamma^{c}$ criteria are chose as $\pm 15 \deg$ and $89 \deg$. This implies that initially a shower is considered active for a full month, which is not true for many showers. This is why the most plausible shower parent is derived considering $\Delta \lambda_{\odot}$. Another issue is that the duration of a shower might change from one year to another, but this is not documented in the IAU meteor shower database.

Because impacts are single points (as opposed to meteors for which a whole trajectory can be computed) it is impossible to identify with 100\% certainty the parenthood of a shower for a given impact. However, our approach allows one to dismiss all showers for which the parenthood is impossible, from geometrical considerations. This is a necessary but not sufficient condition.

In our sample we find that the impacts are associated with 23 meteoroid streams as shown in Table~\ref{table2}. Only event 32 cannot be linked to a confirmed stream and thus is assigned as a sporadic event.

\subsection{Flash temperatures estimation}
\label{temperatures}

Following previous observational and laboratory studies, the flash is assumed to be a black body radiator, whose spectral energy distribution is represented by the Planck formula: 

\begin{equation}
B(\lambda,T) = \frac{2hc{^2}}{\lambda^{5}}\frac{1}{\exp(\frac{hc}{\lambda k_{B}T}) - 1}
\label{eq2}
\end{equation}
calculated in erg cm$^{-2}$ s$^{-1}$ A$^{-1}$ sr$^{-1}$ where $h=6.62\times10^{-27}$~g~cm$^{2}$s$^{-1}$ the Planck constant, $c=3\times10^{10}$~cm~s$^{-1}$ the speed of light, $k_{B}=1.38\times10^{-16}$~g~cm$^{2}$s$^{-2}$K$^{-1}$ the Boltzmann constant, $T$ and $\lambda$ the temperature of the flash and the wavelength, respectively. In this work we also consider that the observer on Earth observes the flux of half black body-like sphere and thus:
\begin{equation}
F =  \frac{\pi r^2}{D^2} B(\lambda, T)
\label{eq3}
\end{equation}
where $r$ (in m) the radius of the black body and $D$ the Earth-Moon distance at the time of the impact.

Since the NELIOTA magnitudes are calibrated in the Cousin system \citep{xilouris2018} the formulas of \cite{bessell1998} can be used to calculate the spectral flux density of each flash and are defined as:
\begin{equation}
\xi_R = \frac {\int_{}^{} B(\lambda,T)R_c(\lambda) d\lambda}{\int_{}^{} R_c(\lambda) d\lambda} 
\label{eq4}
\end{equation}
\begin{equation}
\xi_I =  \frac{ \int_{}^{} B(\lambda,T)I_c(\lambda) d\lambda}{\int_{}^{} I_c(\lambda) d\lambda} 
\label{eq5}
\end{equation}
where $R_c(\lambda)$ and $I_c(\lambda)$ the Cousin filters response. The importance here is that we have to use the formulas above and not directly the isophotal flux at the $\lambda_{eff}$ \cite[i.e. the formulas from Table A.2 of ][]{bessell1998} because the spectral energy distribution of a flash could be (and it is) different than that of an A0 star (flash temperatures are lower). In the previous work of \cite{bonanos2018}, in order to calculate the flux of the black body the theoretical $\lambda_{eff}$ of the Cousin filters as $\lambda_R=641$~nm and $\lambda_I=798$~nm are used and colour corrections due to the difference between the flash temperature and that of the A0 star are neglected. 

\begin{figure}
\includegraphics[width=\columnwidth]{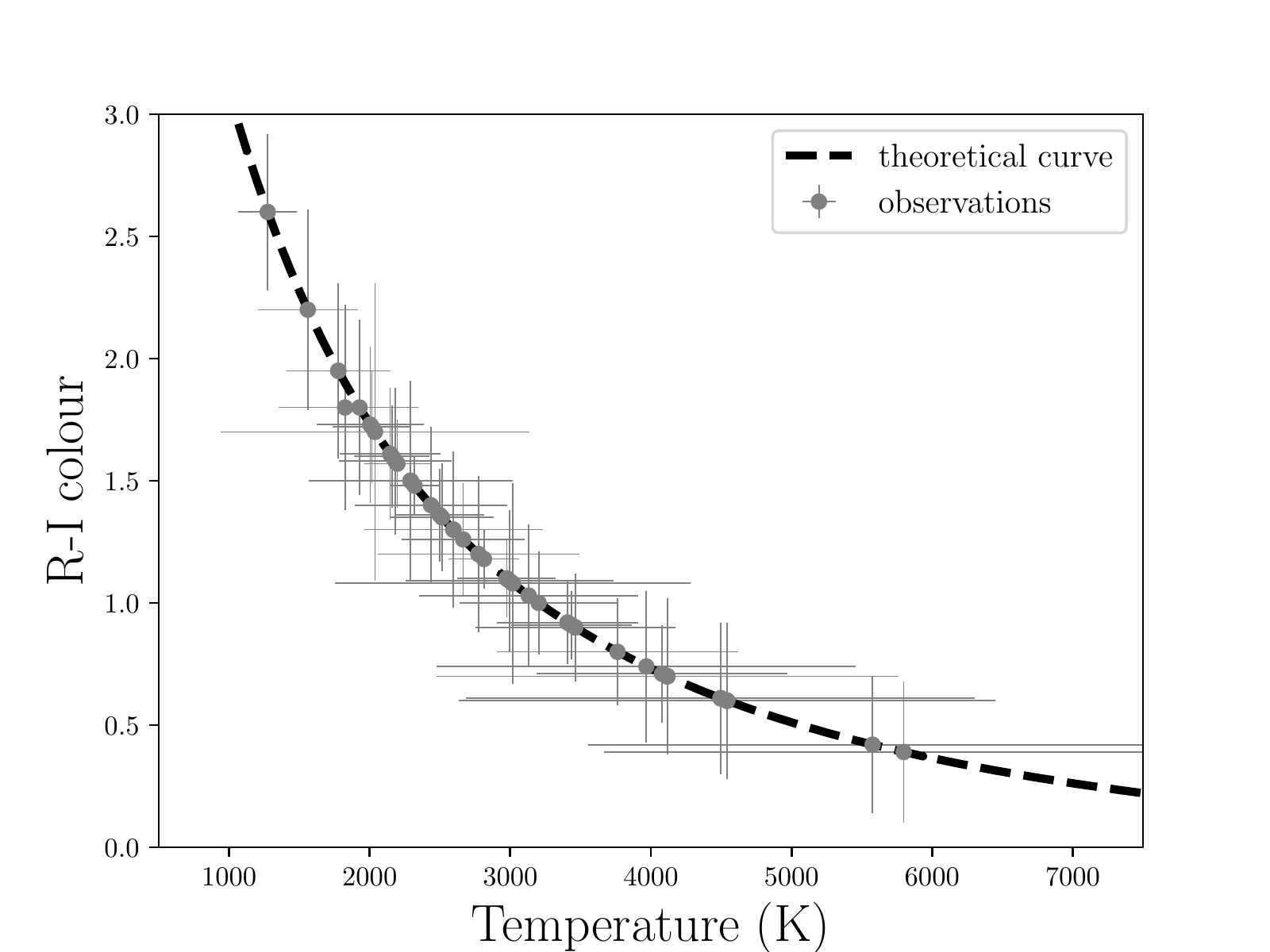}
\caption{Relation of the flash colour index with temperature, assuming a black body radiation.}
\label{fig1}
\end{figure}

The $R$ and $I$ magnitudes and the $R-I$ colour are defined as:
\begin{equation}
R =  -2.5 log(\frac {\pi r^2}{D^2}) -2.5 log (\xi_R) -21.1 -ZP_R
\label{eq6}
\end{equation}
\begin{equation}
I =  -2.5 log(\frac {\pi r^2}{D^2}) -2.5 log(\xi_I) -21.1 -ZP_I
\label{eq7}
\end{equation}
\begin{equation}
R-I =  -2.5 log(\frac{\xi_R}{\xi_I}) -(ZP_R-ZP_I)
\label{eq8}
\end{equation}
where $ZP_R=0.555$ and $ZP_I=1.271$ as defined by \cite{bessell1998}. The value of $R-I$ clearly depends only on $T$, as the rest of terms are simplified in particular the absolute intensity of the black body radiator.
We evaluated the $R-I$ between 500--10,000~K and produced the theoretical curve as function of temperature (Fig.~\ref{fig1}). For each flash, given the $R-I$ from Table~\ref{table1} we obtain its temperature and we can evaluate the values of $\xi_R$ and $\xi_I$.
To estimate the error in $T$ a Monte Carlo approach is adopted with typically $10^{5}$ iterations. At each iteration a temperature value is derived from a new set of $R$ and $I$ which are extracted from Gaussian distributions centred on the nominal $R$ and $I$ values of the flash and standard deviations equal to the magnitude errors given by NELIOTA. The result of the temperature distribution can be approximated by a Gaussian and we consider its standard deviation as the temperature uncertainty. All results are summarised in Table~\ref{table2}.

\subsection{Masses of the impactors}
\label{masses}

The masses of the impactors can now be calculated from its $KE$ using Eq.~\ref{eq1}. The $E_{lum}$ of each event is computed from the luminosity $L$ as: 
\begin{equation}
L = \pi^2 r^2 \int_{400}^{900}B(\lambda,T) d\lambda
\label{eq9}
\end{equation}
\begin{equation}
E_{lum} = L t
\label{eq10}
\end{equation}
where $T$ the temperature as derived from the previous step, $t$ the exposure time of the observations and $B(\lambda, T)$ the Planck integrated over the visible wavelengths (400-900~nm) according to the definition of the $E_{lum}$. The scaling factor here is the radius of the black body $r$, which is calculated from Eq.~\ref{eq6} or Eq.~\ref{eq7} for a given $T$. 

Equation~\ref{eq10} is evaluated for each set of frames that the temperature can be calculated and then is summed to give the total $E_{lum}$. For the events that last one frame it is assumed that the $E_{lum}$ is released over the time of this frame, i.e. 0.033~s. 
In order to obtain masses from Eq.~\ref{eq1} we have:
\begin{equation}
m = \frac{2 E_{lum}}{v_{im}^2 \eta}
\label{eq11}
\end{equation}
where $v_{im}$ are the impact speeds of Table~\ref{table2}. For the events that originated from meteor streams, the $v_{im}$ used is that of the stream $v_{stream}$ corrected to lunar velocity $v_{M}$ which our method of section \ref{origin} provides directly. Regarding the sporadic events, there are several estimations in the literature about the average impact velocity, such as 16~km~s$^{-1}$ \citep{steel1996}, 19.2~km~s$^{-1}$ \citep{ivanov2006}, and 24~km~s$^{-1}$ \citep{mcnamara2004}. Here the mass estimation was done 24~km~s$^{-1}$.

There are several studies that estimated the luminous efficiency, as for example \cite{bellotrubio2000} where $5\times10^{-4}<\eta<5\times10^{-4}$ with a preferred value of $2\times10^{-3}$, while \cite{moser2011} estimated for three different streams $\eta$ to range between $1.2\times10^{-3}$ and $1.6\times10^{-3}$. 
The masses presented in Table~\ref{table2} were calculated using two different values for luminous efficiency, $\eta~=~5\times10^{-4}$ and $\eta~=~1.5\times10^{-3}$ following also the example of \cite{bouley2012}. Note that masses scale inversely proportional to $\eta$ and therefore for a smaller luminous efficiency, e.g. $\eta~=~5\times10^{-4}$, the values of masses from Table~\ref{table2} must be multiplied by 10. For the events when the temperature can be measured in at least two consequent frames the $E_{lum}$ is calculated for each frame-step and is summed. For the events with ID~2, 20, 40, 42, 47, 51 the flash is recorded in more $I-band$ than $R-band$ frames, but for those frames the $T$ measurement is not possible. What is calculated instead is the fraction of the $E_{lum}$ that should remain from the previous time-step. This is different for each flash, implying also different cooling rates. In that way it is possible to sum the energy and have a more accurate value for the mass. However, this correction is not applied for the flashes with ID~13 and 19. In the first case it is uncertain that the flux belongs to one event and as it is described in Section~\ref{cooling} it may be a multiple impact event of several impactors. For the event 19, the photometry of the $R_b$ frame is very uncertain as the flux appears to be in only one pixel making the $E_{lum}$ estimation for the second frame debatable. Thus, lacking a good estimation of the $E_{lum}$ degradation it is preferred not to proceed with any correction.

\ca{Here we have to remind that the method described in section \ref{origin}, for the investigation of the meteoroid parenthood, is necessary but not totally sufficient. Therefore, the calculated impact speeds can be in some cases overestimated leading to underestimated masses.}

\begin{table*}
\centering
\caption{Temperatures, KE and masses calculated using $\eta_1=5\times10^{-4}$ and $\eta_2=1.5\times10^{-3}$ respectively. Note that the mass of the flash with ID~13 is estimated for the total flux and not for each individual peak (see \ref{cooling}) which in the case of a double impact is overestimated.}
\label{table2}
\begin{tabular}{rllc|rrrr}
\hline
\hline
&&&&&$\eta_1=5\times10^{-4}$&&$\eta_2=1.5\times10^{-3}$\\
Flash ID& Shower ID & $T \pm \sigma_T$ (K) & $V_{im}$ (km~s$^{-1}$)  & KE ($\times10^{7}$J) & $m \pm \sigma_m$ (g) &  KE ($\times10^{6}$J) & $m \pm \sigma_m$ (g)\\
\hline
1. 	&AAN	&	2975 $\pm$ 350	&	43.22	&	2.33 	& 24.96 	& 7.77 	& 8.32\\
2.	&XHE	&	4540 $\pm$ 560	&	35.40	&	58.83& 938.78 	& 196.07 	& 312.93\\
3.	&XHE	&	3410 $\pm$ 500	&	35.40	&	5.52	& 88.00 	& 18.38	& 29.33\\
4.	&EVI		&	4080 $\pm$ 890	&	30.22	&	3.98	& 86.97 	& 13.24	& 28.99\\   
5.	&KSE	&	2200 $\pm$ 240	&	48.72	&	2.60	& 21.89 	& 8.66 	& 7.30\\  
6.	&ETA	&	2510 $\pm$ 370	&	64.45	&	2.38	& 11.48 	& 7.94 	& 3.83\\
7.	&NZC	&	1930 $\pm$ 420	&	40.74	&	1.34	& 16.19 	& 4.48 	& 5.40\\
8.	&NZC	&	3020 $\pm$ 1260	&	40.51	&	1.60	& 19.50 	& 5.33 	& 6.50\\
9.	&CAP	&	1825 $\pm$ 470	&	23.57	&	1.18	& 42.67 	& 3.95 	& 14.22\\
10.	&PER	&	1780 $\pm$ 370	&	57.02	&	1.38	& 8.51 	& 4.61 	& 2.84\\ 
11.	&PER	&	2300 $\pm$ 740	&	57.03	&	1.49	& 9.15 	& 4.96 	& 3.05\\ 
12.	&PER	&	3130 $\pm$ 780	&	57.03	&	1.31	& 8.09 	& 4.39 	& 2.70\\  
13.	&PER	&	1780 $\pm$ 140	&	57.03	&	29.56& 181.79	& 98.54 	& 60.60\\
14.	&PAU	&	4500 $\pm$ 1310	&	40.26	&	2.20	& 27.11 	& 7.32 	& 9.04\\
15.	&KCG	&	2180 $\pm$ 400	&	24.86	&	1.65	& 53.35 	& 5.49 	& 17.78\\
16.	&CAP	&	2000 $\pm$ 380	&	19.88	&	1.60	& 81.06 	& 5.34 	& 27.02\\
17.	&ERI		&	2020 $\pm$ 280	&	62.82	&	1.42	& 7.20 	& 4.74 	& 2.40\\
18.	&PER	&	2500 $\pm$ 315	&	59.47	&	2.33	& 13.18	& 7.77 	& 4.39\\
19.	&SPE	&	2975 $\pm$ 230	&	63.34	&	7.08	& 35.28 	& 23.59 	& 11.76\\
20.	&SPE	&	2320 $\pm$ 175	&	62.53	&	17.99& 92.04 	& 59.98 	& 30.68\\
21.	&ORI	&	3430 $\pm$ 430	&	69.13	&	4.96	& 20.78 	& 16.55 	& 6.93\\
22.	&ORI	&	5580 $\pm$ 2025	&	68.98	&	2.05	& 8.63 	& 6.85 	& 2.88\\
23.	&ORI	&	2670 $\pm$ 440	&	67.56	&	1.49	& 6.55 	& 4.98 	& 2.18\\
24.	&ORI	&	4500 $\pm$ 1810	&	65.45	&	2.94	& 13.75 	& 9.81 	& 4.58\\
25.	&NTA	&	3200 $\pm$ 560	&	24.92	&	2.13	& 68.64 	& 7.10 	& 22.88\\
26.	&LEO	&	5800 $\pm$ 2130	&	70.23	&	2.15	& 8.74 	& 7.18 	& 2.91\\
27.	&GEM	&	2150 $\pm$ 360	& 	34.16	&	2.26 	& 38.79 	& 7.54 	& 12.93\\
28.	&GEM	&	3760 $\pm$ 1490	&	34.16	&	1.65	& 28.32 	& 5.51 	& 9.44\\
29.	&GEM	&	4540 $\pm$ 1560	&	33.85	&	1.75	& 30.50 	& 5.82 	& 10.17\\
30.	&GEM	&	2974 $\pm$ 250	&	33.56	&	20.72 & 367.88	& 69.06 	& 122.63\\
31.	&NCC	&	4117 $\pm$ 930	&	25.95	&	2.96	& 88.05 	& 9.88 	& 29.35\\
32.	&	-	&	2595 $\pm$ 242	&	24.00        &	3.06	& 106.23 	& 10.20	& 35.41\\
33.	&KSE	&	4120 $\pm$ 595	&	45.06	&	8.90	& 87.69 	& 29.67 	& 29.23\\
34.	&SSG	&	3460 $\pm$ 700	&	25.79	&	2.93	& 88.05 	& 9.76 	& 29.35\\	
35.	&NZC	&	3460 $\pm$ 710	&	44.04	&	2.81	& 28.97 	& 9.36 	& 9.66\\	
36.	&ARI		&	4542 $\pm$ 1910	&	44.08	&	2.85	& 29.31 	& 9.49 	& 9.77\\	
37.	&ARI		&	2040 $\pm$ 1100	&	44.08	&	2.58	& 26.57 	& 8.60 	& 8.86\\	
38.	&SZC	&	1275 $\pm$ 210	&	37.59	&	2.06	& 29.14 	& 6.86 	& 9.71\\	
39.	&PAU	&	2160 $\pm$ 270	&	43.65	&	4.21	& 44.21 	& 14.04 	& 14.74\\	
40.	&PER	&	2290 $\pm$ 155	&	58.40	&	7.69	& 45.10 	& 21.26 	& 15.03\\	
41.	&SDA	&	2290 $\pm$ 490	&	36.68	&	1.42	& 21.13 	& 4.74 	& 7.04\\	
42.	&PER	&	2980 $\pm$ 240	&	58.35	&	10.03& 58.95 	& 33.45 	& 19.65\\	
43.	&PER	&	4540 $\pm$ 1340	&	58.35	&	2.29	& 13.43 	& 7.62 	& 4.48\\	
44.	&SDA	&	4540 $\pm$ 1890	&	36.65	&	1.74	& 25.85 	& 5.77 	& 8.62\\	
45.	&SDA	&	2770 $\pm$ 720	&	36.70	&	0.93	& 13.84 	& 3.11 	& 4.61\\	
46.	&PER	&	4120 $\pm$ 1045	&	58.28	&	0.88	& 5.16 	& 2.92 	& 1.72\\	
47.	&PER	&	2970 $\pm$ 250	&	58.28	&	12.98& 76.44 	& 43.27 	& 25.48\\	
48.	&SDA	&	2440 $\pm$ 540	&	36.69	&	1.00	& 14.95 	& 3.35 	& 4.98\\	
49.	&SDA	&	1560 $\pm$ 360	&	36.91	&	0.91	& 13.38 	& 3.04 	& 4.46\\	
50.	&NIA		&	4120 $\pm$ 1640	&	27.25	&	2.75	& 74.02 	& 9.16 	& 24.67\\
51.	&NIA		&	2770 $\pm$ 220	&	27.16	&	47.45& 1286.44& 158.16	& 428.81\\
52.	&NIA		&	2290 $\pm$ 725	&	27.15	&	1.70	& 46.13 	& 5.67 	& 15.38\\
53. 	& NUE      &       4120 $\pm$ 1620	&	71.35        &	0.96	& 3.78	& 3.21 	& 1.26\\
54. 	&  NUE     &       2600 $\pm$ 635	&	71.34	&	0.88	& 3.47 	& 2.94 	& 1.16\\
55. 	& ORI      	&       3760 $\pm$ 860	&	68.86    	&	4.24 	& 17.90 	& 14.15 	& 5.97\\
\hline
\hline
\end{tabular}
\end{table*}

\subsection{Sizes of the impactors and prediction for the produced craters}
\label{sizes}

\begin{figure}
\includegraphics[width=\columnwidth]{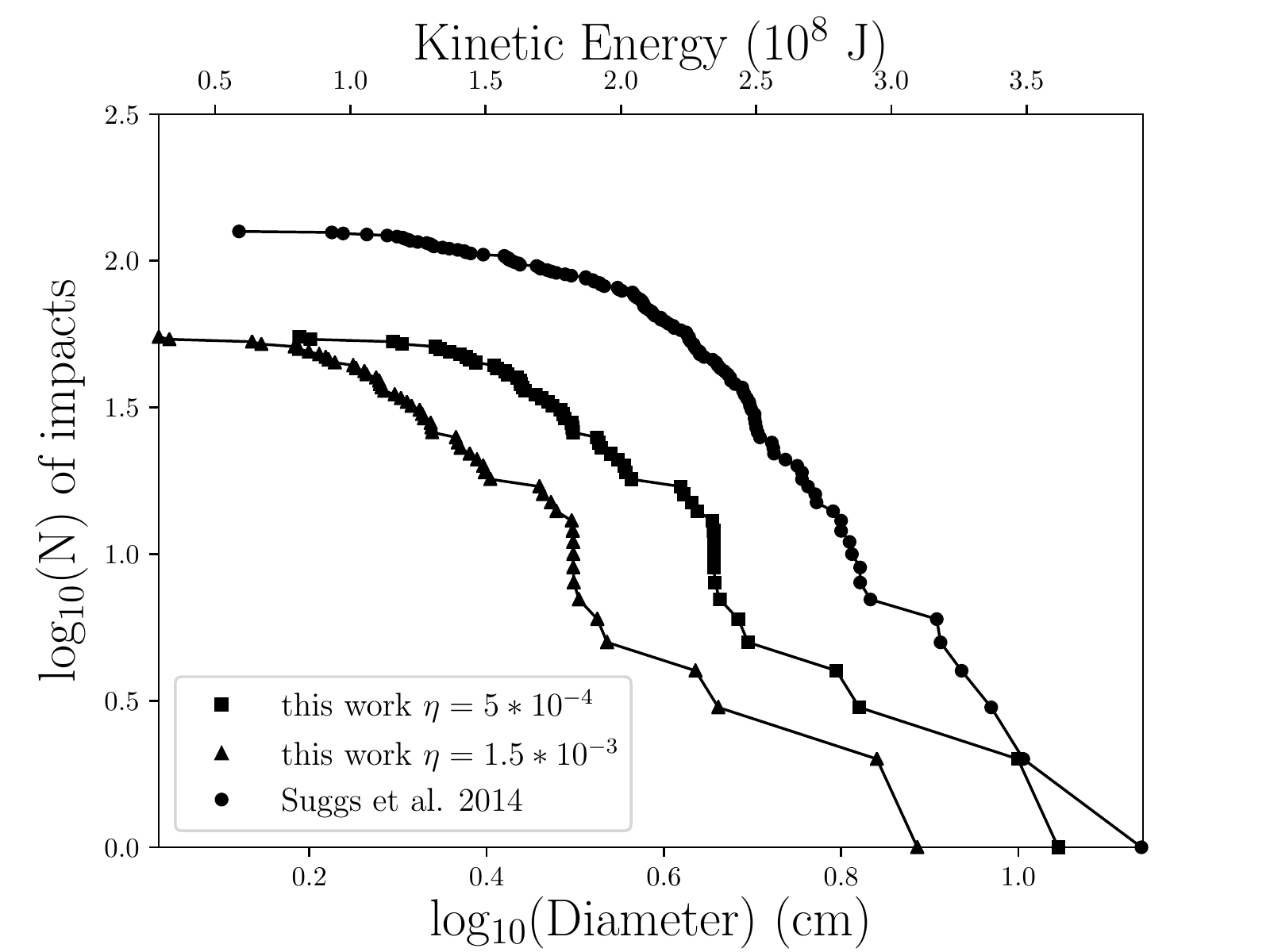}
\caption{ The SFD of NELIOTA observations presented in log(D) vs. log(N) space, is calculated with two different $\eta$ with a typical error of $\sim30\%$. If $\eta$ is changed the SFD is shifted along the x-axis but the slope is unchanged. Both samples, from \protect\cite{suggs2014} and this work, include events linked to meteoroid streams and sporadic population.}
\label{fig2}
\end{figure}

Having a measurement for the impacting mass and assuming spherical-shaped objects it is also possible to estimate their size (Fig.~\ref{fig2}). The important parameter in this calculation is the choice of the impactor's density. In principle for meteoroids there are two estimations for density. The first is the mineralogical density, $\delta_{m}$, that is related to the specific material of the objects and the other is the bulk density, $\delta$, which has lower values since it is taken into account the porosity of the object.

In this work we try to assign a bulk density to the impactors, that is compatible with the parent body of their stream, and is taken from \cite{babadzhanov2009}. Orionids and Eta Aquarids are associated with comet Halley and thus is used the same $\delta$ value for both meteoroid populations. For the sporadic and any other event that has no density estimations is used $\delta=1,800$~kg~m$^{-3}$ \citep{babadzhanov2009} as an average value derived from observations (Table~\ref{table3}). The actual bulk density for a single sporadic meteoroid may vary from low values to higher values for the dark carbonaceous and stony/metallic parents respectively.

\begin{table}
\centering
\caption{Estimated impactor size ranges, using meteoroid bulk densities from \protect\cite{babadzhanov2009} and sizes of the potential craters (using $\eta=1.5\times10^{-3}$). The sizes of impactors and their potential craters have double values when an order of magnitude smaller $\eta$ is adopted.} 
\label{table3}
\begin{tabular}{lc|ccc}
\hline
\hline
 Origin 	&  density		&	 $d_{im}$ 	\\
		&(g~cm$^{-3}$)&		(cm)	 \\
\hline
GEM	&	2.9		&	1.9--4.3\\
PER 	&	1.2		&	1.4--4.6\\
ORI	&	0.9		&	1.7--2.5\\
NTA	&	1.6		&	3.0\\
ETA	&	0.9		&	2.0\\	
KCG	&	2.2		&	2.5\\
SDA&	2.4		& 1.5--1.9\\
LEO&	0.4		&  2.4 \\
CAP&	2.1		&  2.3--2.9\\
SPO\&others	&1.8	&1.0--7.7\\
\hline
\hline
\end{tabular}
\end{table}

\subsection{Temperature cooling}
\label{cooling}

\begin{figure}
\includegraphics[width=\columnwidth]{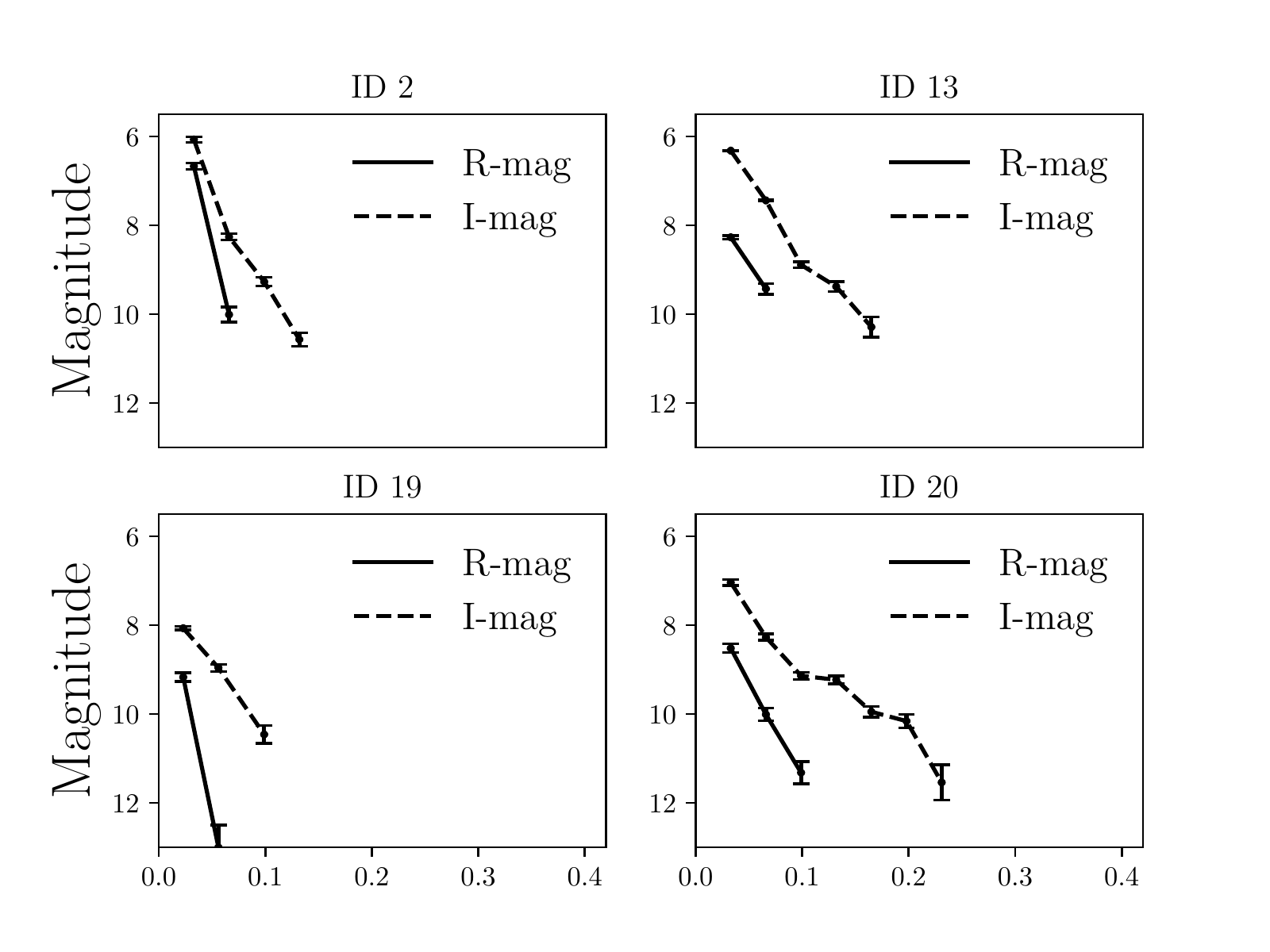}
\vspace{-10pt}
\includegraphics[width=\columnwidth]{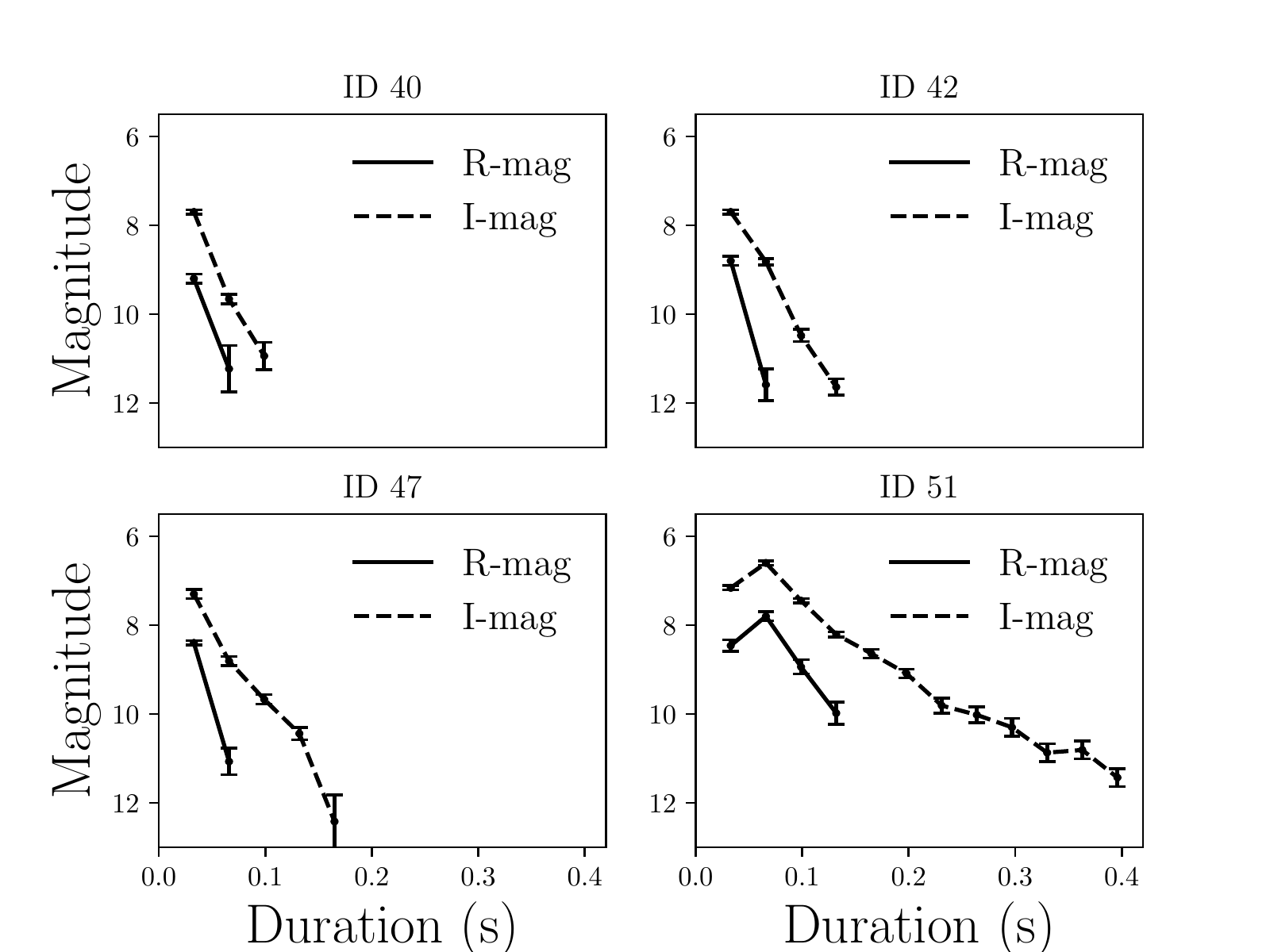}
\includegraphics[width=\columnwidth]{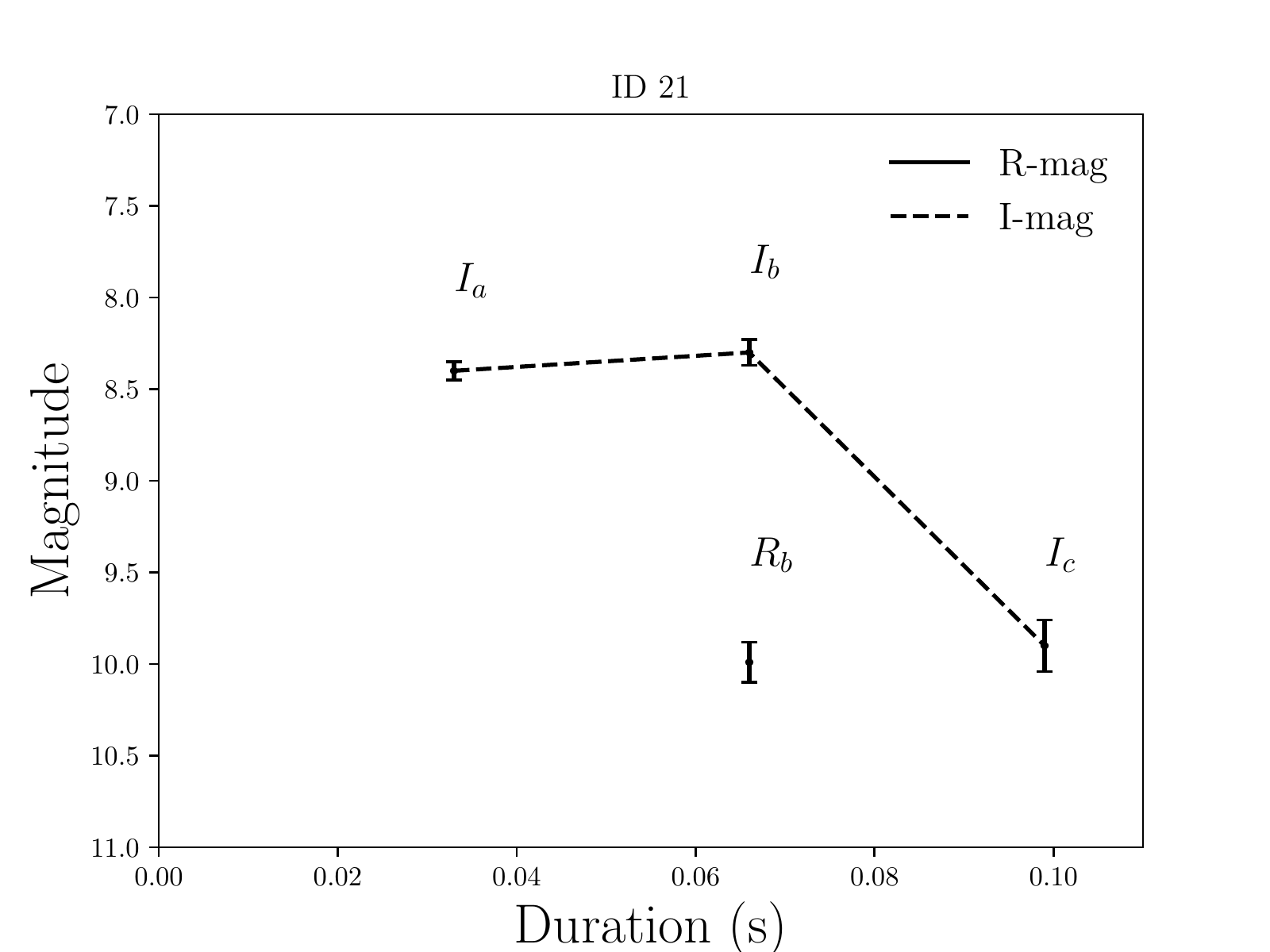}
\caption{Magnitude drop for the events that were detected in more than two frames.}
\label{fig3}
\end{figure}

\begin{figure}
\includegraphics[width=\columnwidth]{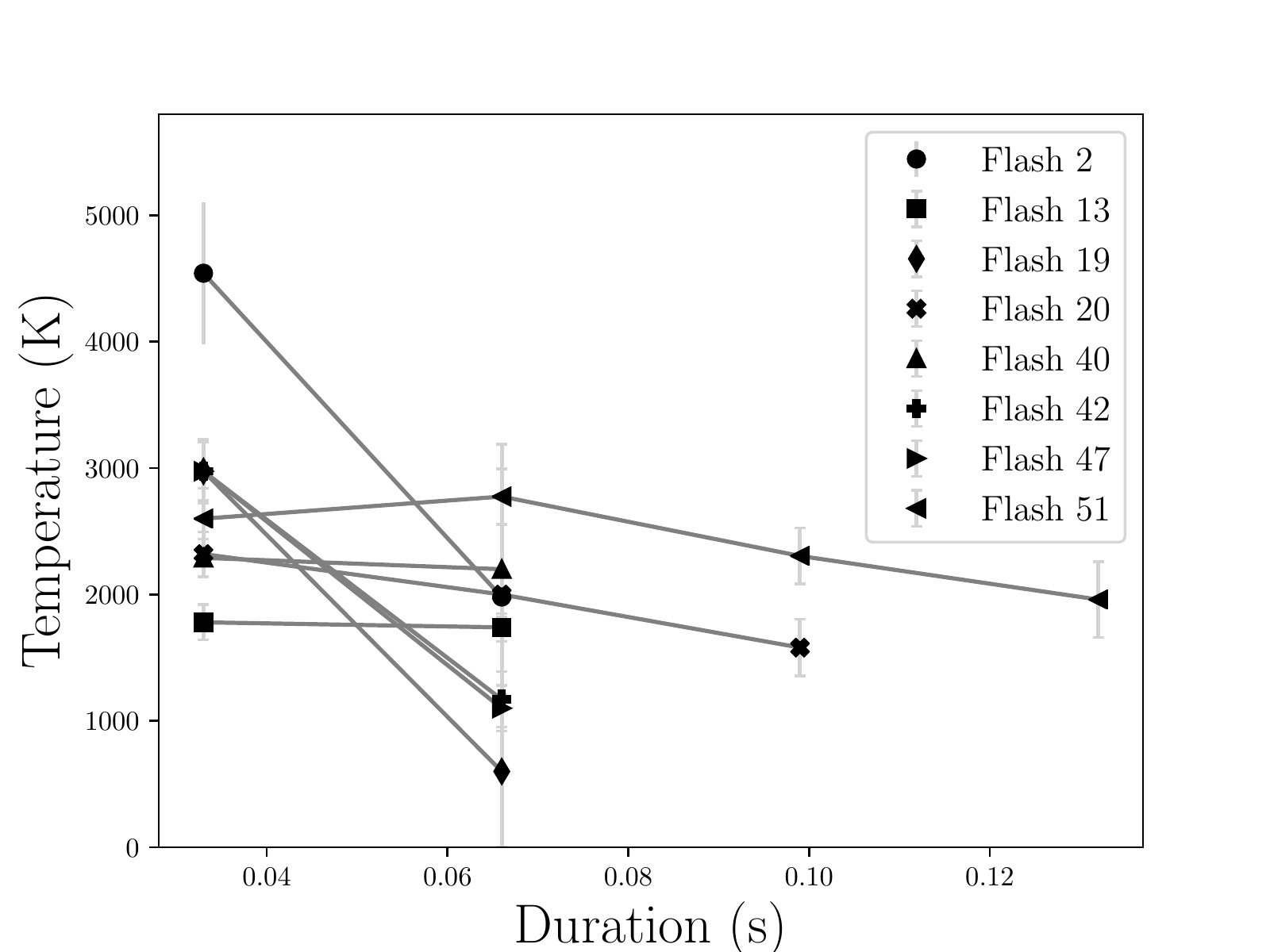}
\caption{Cooling of the ejected material after the impact event. Each flash appears to have different cooling rate. The event with ID~13 appears to have initially constant $T$ which could be an indication of a slightly non-synchronous double impact.}
\label{fig4}
\end{figure}

The impact flashes are rapid events, however a few of them last long enough to appear in several frames. This allows the temperature estimation at several stages during the cooling process. This cooling time in turn will help towards the understanding of the process or processes that produce the impact flash. In the current sample there are four events that lasted at least two frames in both filters (see Fig.~\ref{fig3} and Table~\ref{table4}), and thus the $T$ evolution with time can be measured (Fig.~\ref{fig4}). The rest of the events have recorded flux in one band in two or more consequent frames but only in one frame at the other band. This means that $T$ cannot be estimated for more than one point during the total duration of the flash, the cooling is not time resolved, and therefore is not presented here.

There are two interesting events with ID~13 and ID~21, not showing similar pattern as all the others.
During the event with ID~21 it appears that the flash has almost constant magnitude in $I-band$ (if not brighter) and then fades again. Specifically the flash can be detected in three frames in the $I-band$ ($I_a$, $I_b$, $I_c$), with $I_a \sim I_b$ within the error, but surprisingly in the R-band the flash is only detected in $R_b$, even after a visual inspection. This observed fluctuation in magnitude could be due to the unique capture of the heating and cooling of the ejected material during the impact. Considering that in laboratory experiments we have seen very rapid increase of the flux of the flash both in R and I but much slower cooling, we would probably expect not to be able to capture this initial part of the event. 

\begin{figure}
\includegraphics[width=\columnwidth]{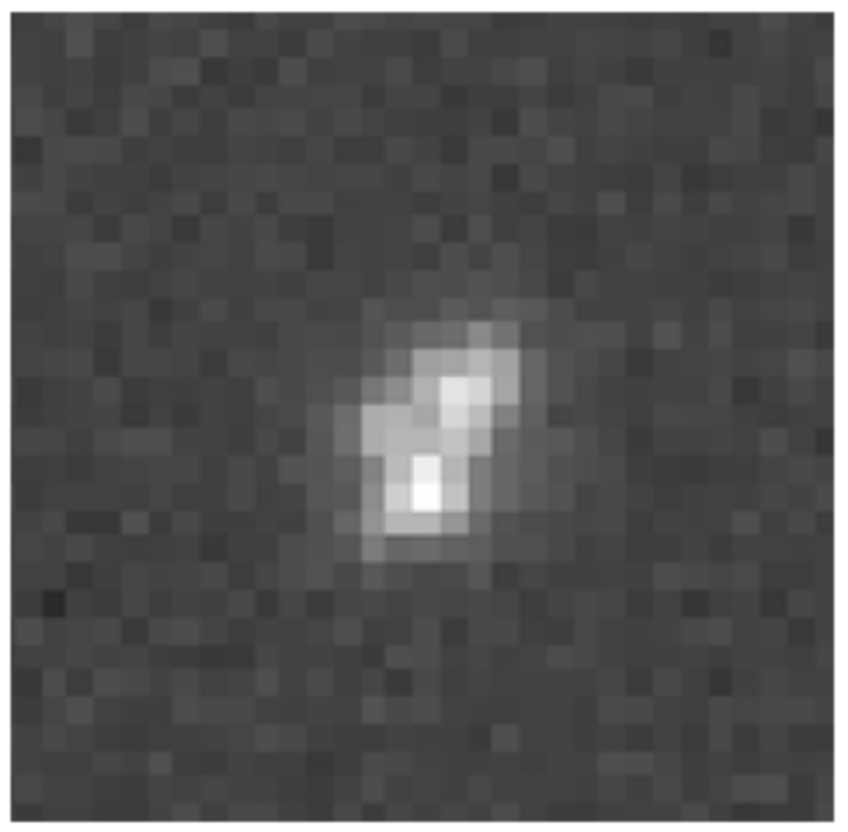}
\caption{The flash with ID~13 in the $I-band$, where are distinguished three peaks.}
\label{fig5}
\end{figure}

The other interesting event is with ID~13. In this particular event, although there is magnitude drop in both bands, the $T$ is constant within the error-bar for the first 0.066~s (two frames). 
By examining the available files of the $I-band$ ($I_a$, $I_b$, $I_c$, $I_d$), the event does not appear to be a single source of flux but has multiple adjacent peaks (see Fig.~\ref{fig5}). The total formation is elongated and the same orientation of the elongation is noticed in the $R-band$ files ($R_a$, $R_b$). However, in the later no separate peaks are distinguished and this fact makes difficult the temperature calculation for each pair of the relevant peaks as before. In general, rapid variations in the atmosphere can cause this effect of separating the source light in multiple sources (speckles). In this particular case the event lasts at least 0.132~s which is quite long but still possible to be atmospheric speckles (J. Drummond, private communication). 
The camera's pixel scale is 0.4~arcsec/pixel. The distance between the centre of light of the two largest peaks is 3.87~pixels, i.e. 1.55~arcsec. A field containing a magnitude calibrating star, was observed with several minutes time difference and at similar airmass of the Moon with an exposure time of 2~s. The \ca{PSF} profile of the calibrating star in the $I-band$ is 2.82 and 2.47 pixels for the major and minor axis respectively, corresponding effectively to an area with 5.3~pixels in diameter. The equivalent 
\ca{PSF} profiles in the $R-band$ are 2.68 and 2.4 pixels with an effective area of 5.04 pixels in diameter. By comparing these measurements the identified peaks are inside the PSF of the calibrating star, indicating a speckle.

In the case of a speckle by summing the fluxes of all separated points we get the total flux of the source. Interestingly Fig.~\ref{fig4} shows constant temperature in two subsequent frames which is quite unexpected as typically we expect a rapid temperature drop. The fact that we do not see a temperature drop could be due to multiple impacts. In this scenario it is assumed that an initial small body broke in pieces which impacted within short time difference. In that way while the $T$ of the flash of the first impactor started dropping the second impact occurred keeping the value high, measuring constant $T$ for two subsequent frames. The impact time difference of the objects should be 0.033~s or shorter.
The question that arises is which is the mechanism that forms a small double impactor. In general a single NEO can be disrupted in the atmosphere and the fragments can be separated by differential drag deceleration. However, this mechanism is not applicable in the case of the lunar surface due to lack of an important atmosphere. Another possibility is that the impactor was disrupted by tidal forces as it approaches the target or that was already fragmented in its orbit \citep{chappelow2008}. The later case cannot be explained by collisions since the number of collisions on a small body is proportional to its cross section. The most probable scenario is the fragmentation due to thermal effects, as it has been shown at different scales \citep{delbo2014,granvik2016,pajola2017}. Clusters of meteoroids reaching the Earth's atmosphere have been observed, having very small time difference as for the case of Leonids and the examination of their case gives more evidence of NEO fragmentation in space prior the impact on Earth \citep{watanabe2003}. \ca{The specific case described here is possibly connected to the Perseids meteor stream and further study is needed to associate with a time of perihelion passage.}

\begin{table}
\centering
\caption{Flashes that were recordered in at least two subsequent frames in both filters. Magnitudes and temperatures were calculated as described in \ref{photometry} and \ref{temperatures} respectively. \ca{Temperatures reported earlier are recomputed in this work with the updated method.} $^*$These magnitudes were taken directly from \protect\cite{bonanos2018,xilouris2018}. $^**$These magnitudes were taken from NELIOTA web database.}
\label{table4}
\begin{tabular}{r|crrl}
\hline
\hline
Flash 	&  frame 		& 	$R \pm \sigma_R$ 	& 	$I \pm \sigma_I$	&  $T \pm \sigma_T$ \\
ID		&		 	&	(mag) 			& 	(mag) 			&  (K)\\
\hline
2.	  	&  a$^*$			&	   6.67 $\pm$ 0.07  	&  	 6.07 $\pm$ 0.06 	 & 4540  $\pm$ 560\\
 		&  b$^*$			&	   10.01 $\pm$ 0.17 	&  	 8.26 $\pm$ 0.07 	 & 1980 $\pm$ 210\\
\hline
13.	 	&  a$^*$			& 	   8.27 $\pm$ 0.04 	& 	 6.32 $\pm$ 0.01 	 & 1780 $\pm$  140\\
		&  b$^*$			&	   9.43 $\pm$ 0.12 	&  	 7.44 $\pm$ 0.02 	 & 1740 $\pm$  110\\
\hline
19.	  	& a$^*$			&	  9.17 $\pm$ 0.10	& 	 8.07 $\pm$ 0.04	 & 2975 $\pm$ 230\\
		& b				&	  13.00 $\pm$ 1.50	& 	 8.96 $\pm$ 0.08	 & 600 $\pm$ 600\\
\hline
20.	 	& a$^*$			&	  8.52 $\pm$ 0.10 	& 	7.04 $\pm$ 0.07	& 2320 $\pm$ 175\\ 
  		& b				&	  10.01 $\pm$ 0.14  	& 	8.27 $\pm$ 0.07	 & 2000 $\pm$ 180 \\ 
  		& c				&	  11.32 $\pm$ 0.25 	& 	9.14 $\pm$ 0.08	 &  1580 $\pm$ 225\\ 
\hline
40.		& a$^{**}$			&	9.20$\pm$ 0.10 	& 	7.7$\pm$ 0.05	& 2290 $\pm$ 150\\
		& b				&	11.23$\pm$ 0.52 	& 	9.66$\pm$ 0.11	& 2200 $\pm$ 990\\
\hline
42.		& a$^{**}$			&	8.80$\pm$ 0.10 	& 	7.70 $\pm$ 0.05 	& 	2975 $\pm$ 240\\
		& b				&	11.59$\pm$ 0.36 	& 	 8.82	$\pm$0.07 	& 	1170 $\pm$ 220\\
\hline
47.		& a$^{**}$			&	8.40$\pm$ 0.05 	& 	7.30$\pm$ 0.10 	& 2975 $\pm$  250\\
		& b				&	11.07 $\pm$ 0.30	& 	8.81 $\pm$ 0.10	& 1100 $\pm$  180\\
\hline
51.		& a$^{**}$			&	8.46 $\pm$  0.13	& 	7.16$\pm$0.05	& 	2600$\pm$240\\
		& b				&	7.80 $\pm$  0.10	& 	6.60$\pm$0.05	& 	2775$\pm$220\\
		& c				&	8.94 $\pm$  0.16	& 	7.45$\pm$0.05	& 	2305$\pm$245\\
		& d				&	9.98 $\pm$  0.25	& 	8.21$\pm$0.06	& 	1960$\pm$300\\
\hline
\hline
\end{tabular}
\end{table}

\section{Discussion}
\label{discussion}

\begin{figure}
\includegraphics[width=\columnwidth]{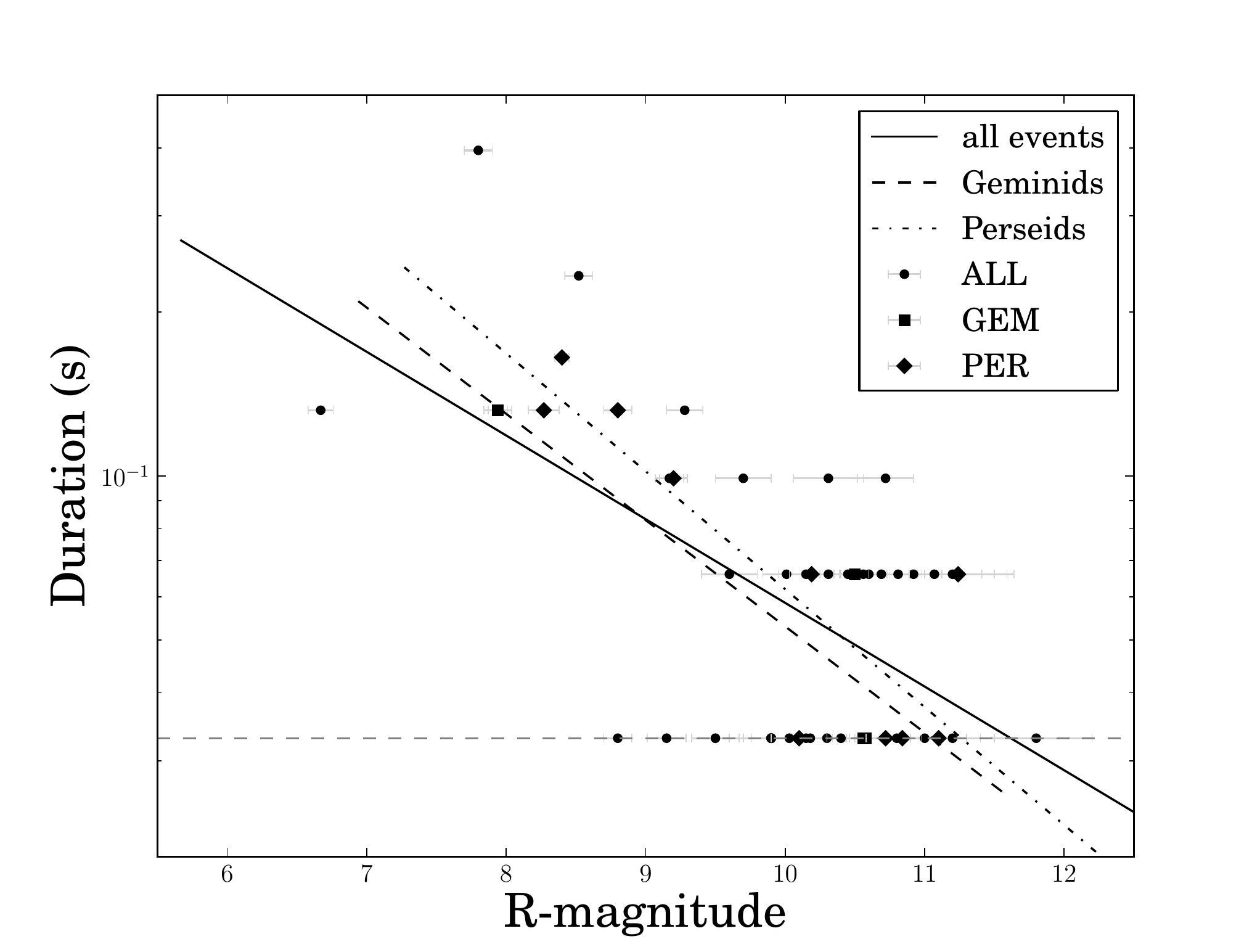}
\includegraphics[width=\columnwidth]{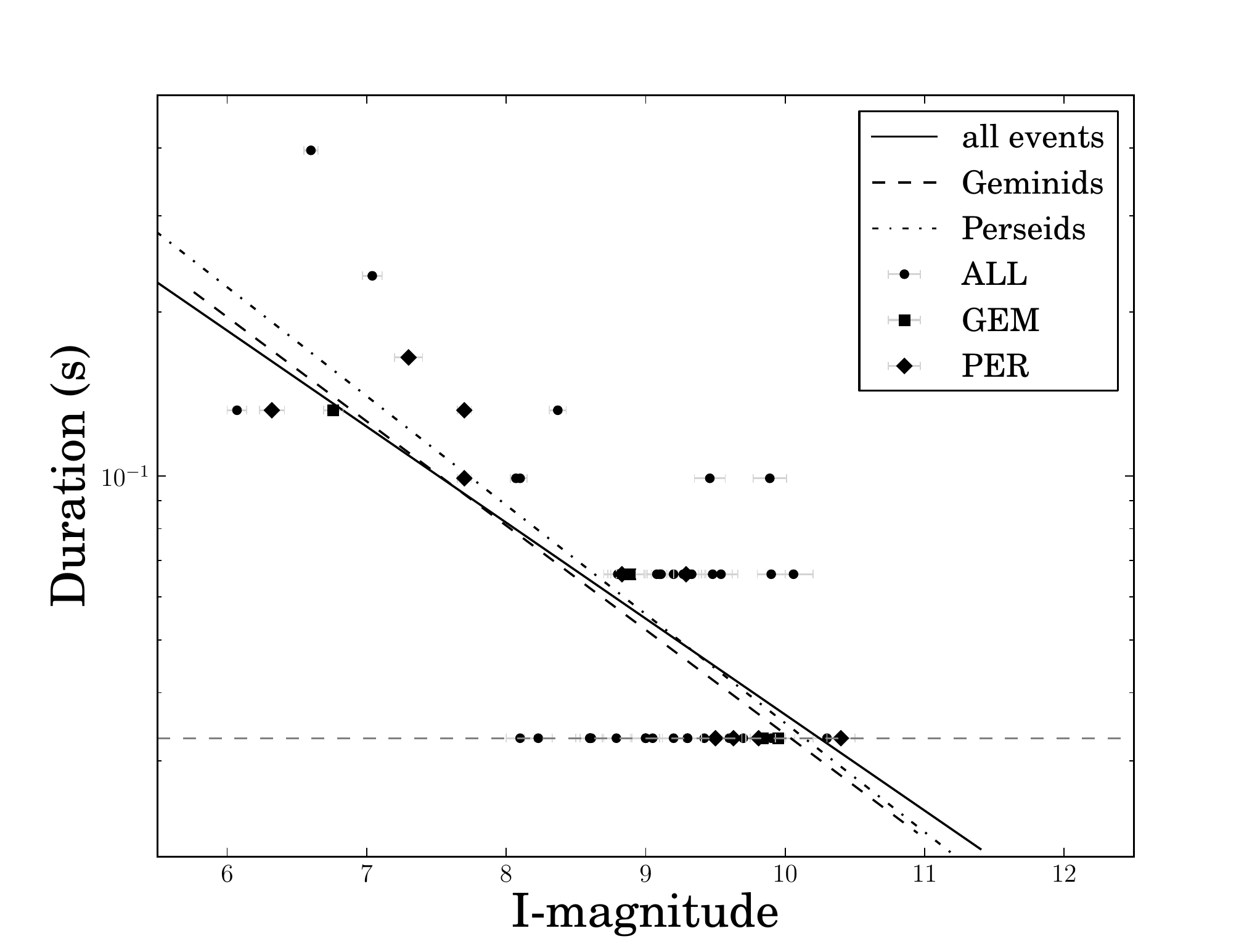}
\caption{Duration of the observed flashes as a function of their $R$ and $I$ magnitudes.}
\label{fig6}
\end{figure}

Here we present lunar impact flash observations with simultaneous detections in $R$ and $I-band$, which allow us to estimate their temperature. Flashes appear brighter in $I-band$, while in $R-band$ have been observed faint events down to almost $12^{th}$ magnitude (Fig.~\ref{fig6}). This is one magnitude fainter than the events that have been presented in previous large surveys \citep{bouley2012, suggs2014}. There is a marginal correlation between the flash duration and the magnitude in both filters. In this sample this correlation appears weaker than the one presented by \cite{bouley2012}. 

\begin{figure}
\includegraphics[width=\columnwidth]{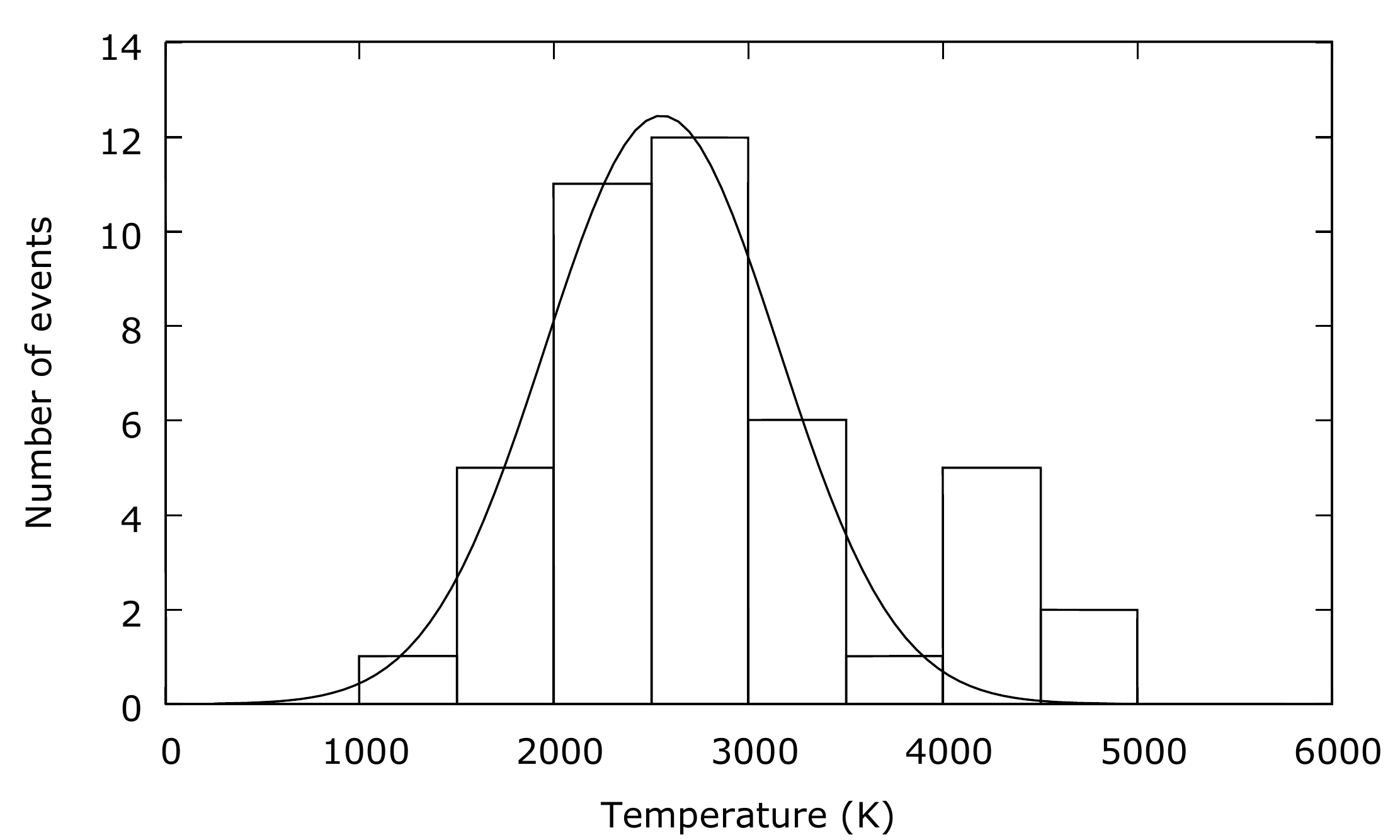}
\caption{Temperature distribution of the observed flashes with a temperature uncertainty less than 30\%. The values correspond to the peak temperature measured from the initial $R$ and $I$ frames of each flash.}
\label{fig7}
\end{figure}

Until very recently, temperatures could not be derived from lunar telescopic observations and thus, when needed, studies \citep{suggs2014,suggs2017} used an average value \citep[e.g. 2,800~K][]{nemtchinov1998}. In Fig.~\ref{fig7} we plot the number distribution of those events with a temperature uncertainly smaller than 30\% (i.e. 43 events), and the best fit Gaussian function of the form $A*exp(-(T-T0)^{2}/2/\sigma^{2})$ for which we find $T_0$ = 2550~K, $\sigma$ = 600 =~K, and A=12.5. This indicates that a good estimate for the typical temperature of the flashed measured by NELIOTA is about 2,500--2,600~K very close to previous assumptions of the average temperature of the lunar flashes.

\begin{figure}
\includegraphics[width=\columnwidth]{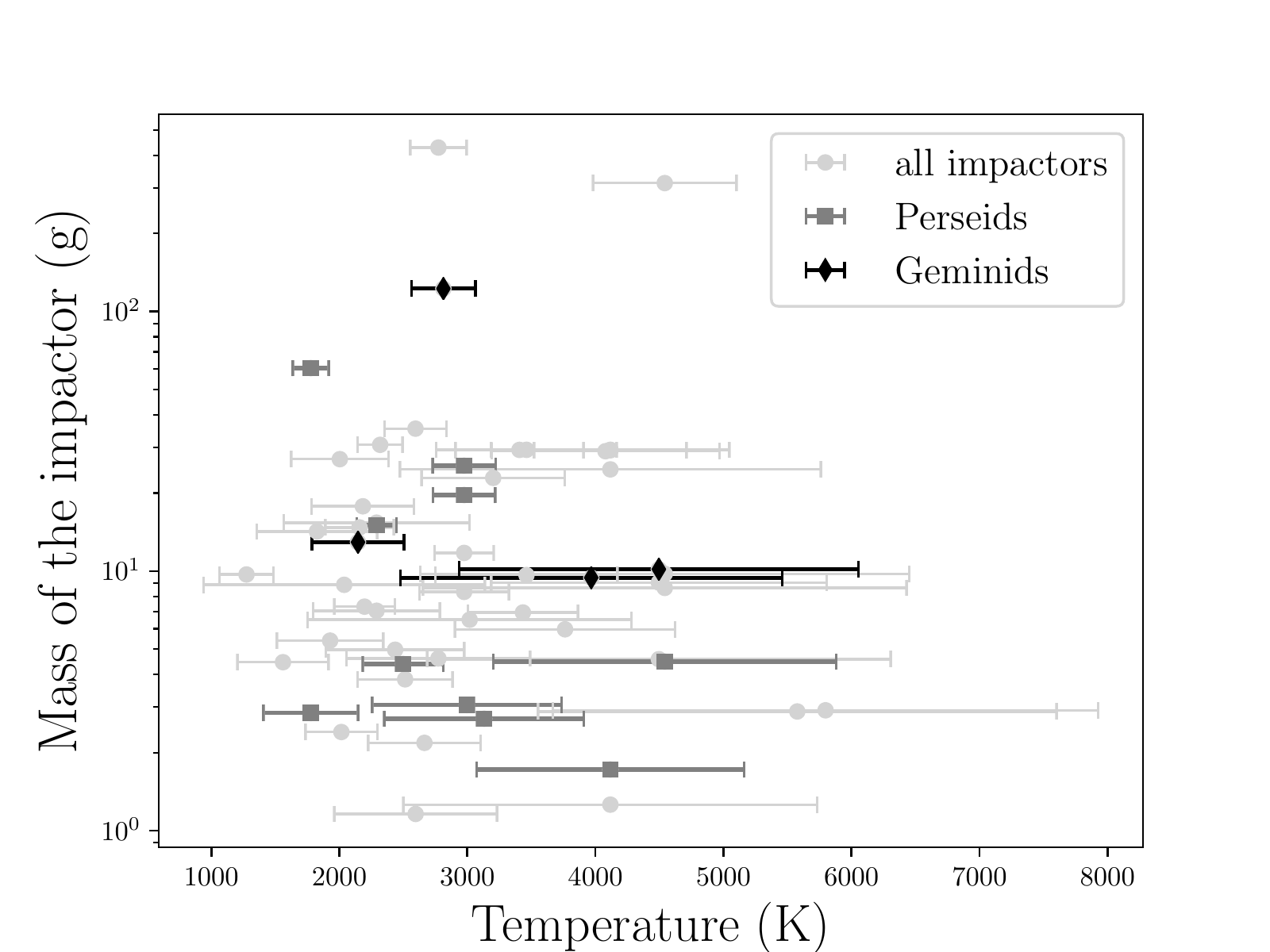}
\caption{For a given impactor mass temperature appears to have a large dispersion in values and therefore there is no obvious correlation. This is because other parameters, such as the type of the material of both of the target and impactor may affect the result.}
\label{fig8}
\end{figure}

Theoretical studies have given us limits for the initial temperatures of the flashes in the range between the melting point of the lunar regolith at 1,725~K and its vaporisation at 3,776~K \citep{cintala1992}. In our sample, 39 out of 55 events are in the aforementioned temperature range, with 15 events having temperatures hotter than 3776~K and only 1 event lower than 1725~K. Indeed, Fig~\ref{fig7} shows a tail of high temperature flashes that deviates significantly from the best fit Gaussian. This observation requires a further investigation to understand the phenomena that produce the light and how big is the deviation from a perfect black body emission. However, regolith temperature before an impact plays a role in the energy partition; it has to be noted that for the aforementioned flash temperature range indicated by \cite{cintala1992} the initial regolith temperature was assumed to be 273~K. As NELIOTA observes flashes on the non illuminated part of the Moon, regolith temperatures are significantly lower than 273~K, It has to be noted that, due to the readout time of the cameras and the moderate integration time, it is possible that the temperatures observed by NELIOTA are not the peak temperature of each individual flash.
After 22 continuous months of NELIOTA observations and using the total sample, there is neither an obvious correlation between maximum temperature and duration of the flash, nor between mass of the impactor and the flash duration. In addition there is no correlation/anti-correlation between the mass of those impactors originating from the same meteoroid stream and the temperature Fig.~\ref{fig8}. A lack of low-mass, low-temperature events is expected for a magnitude-limited survey, such as NELIOTA. 

It has been demonstrated in the laboratory that the intensity of the impact flash, and thus the temperature, is affected by the impact speed and the mass of the impactors. However, all these pioneering studies used dust accelerators in order to achieve high impact speeds of tens of km~s$^{-1}$ \citep{eichhorn1975,eichhorn1976, burchell1996A,burchell1996B}. Moreover, the earliest ones made use mostly of metals and generally materials that are not compatible with the lunar surface and the NEO population. These conditions are important to be taken into account when we try to compare the temperatures of the aforementioned studies with the ones presented here as the materials of both target and impactor have an influence. For example, when several metals were tested, their melting point is anti-correlated with the measured temperature \citep{eichhorn1975}. The temperature range can be different for each experimental campaign, as different masses, compositions and speeds for impactors are used. One example for impact experiments using the van-de-Graaff dust accelerator, impacting iron particles onto several metallic surfaces, is that for impact speeds at 1--20~km~s$^{-1}$ the estimated temperature is 2,000--2,500~K. When gas-gun was used shooting pyrex projectiles on to pumice dust the temperatures were found to be 3,000--4,000~K, for modest impact speeds at about 5.5~km~s$^{-1}$.

In the current sample, the majority of the impactors have low densities because either they come from meteor streams linked to comets or in the case of the sporadic events a low density is adopted. The main exception are the Geminids which originate from the B-type near-Earth asteroid (3200) Phaethon and have bulk densities 2,900~kg~m$^{-3}$ (see Table~\ref{table3}). 

Regarding the SFD of the impactors, we approximate it by a power-law N($>$D) = A*D$^{\alpha}$. The slope $\alpha$ of the power law is independent of the choice of $\eta$. The best fit value of $\alpha$ is -2.28 $\pm$ 0.08, which is in very good agreement with the one observed by \cite{suggs2014} who derived a slope of $\alpha=-2.18 \pm 0.10 $.

\begin{figure}
\includegraphics[width=1.2\columnwidth]{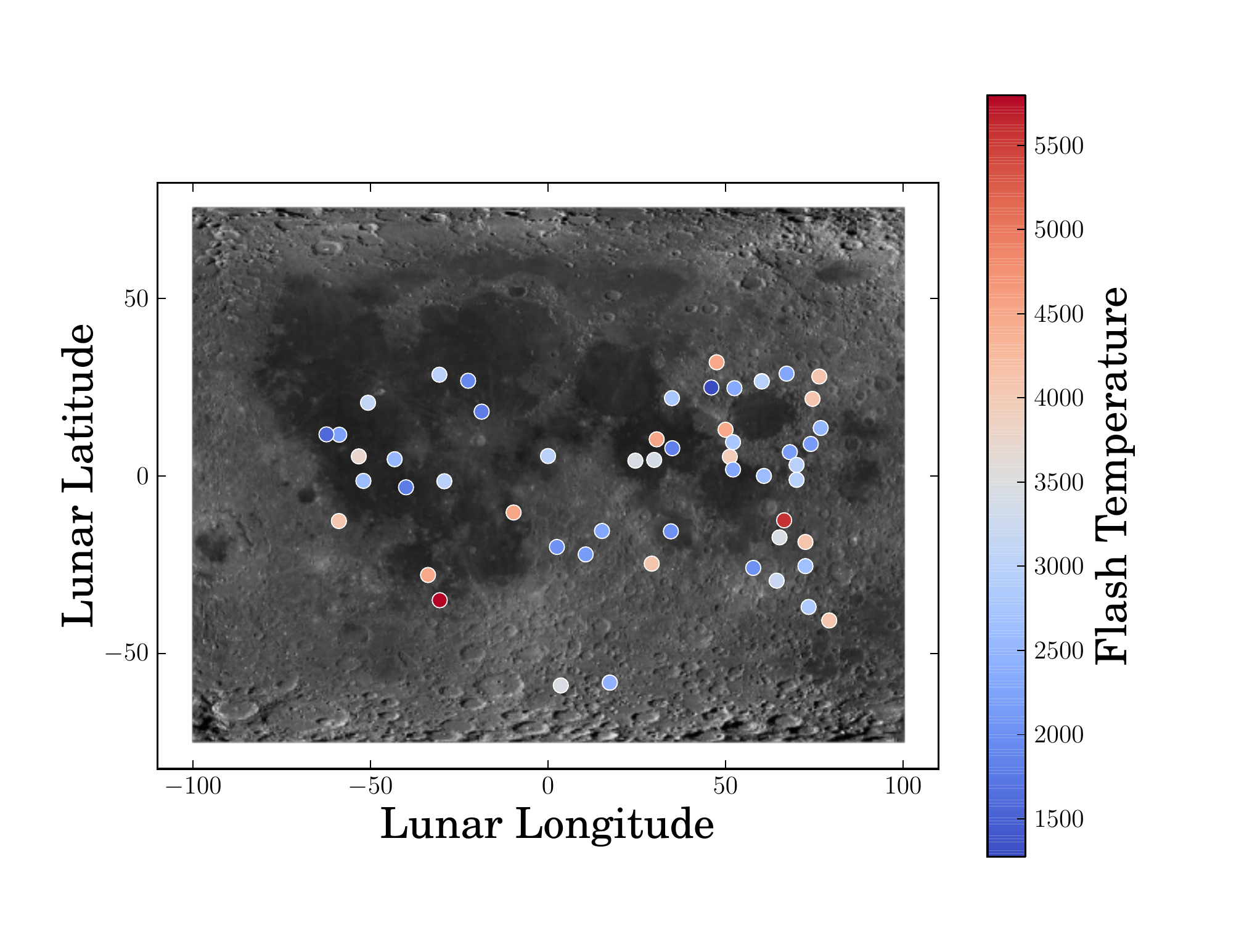}
\caption{Locations of the validated impact flashes along with the measured temperature.}
\label{fig9}
\end{figure}

\section{Conclusions}
\label{conclusions}
In this work is presented the methodology to measure the temperatures of the lunar impact flashes from NELIOTA 2-colour measurements by using the total response of the detectors and not only the peak wavelength as in previous studies. There is an attempt to assign the impacting population to meteoroid streams, providing a more precise estimation of the impact velocity which is an important parameter to calculate the kinetic energy of the impactor. During the impact this initial energy is divided into several parts, with one to be transformed to luminous energy. A great unknown is the energy percentage that is consumed to produce the light. Impactors' masses are presented, while their sizes are calculated considering the documented bulk density of the parent objects. The above conclusions are based on the assumption of the black body radiation. Since the temperatures are directly measured at large scales they can be used for the estimation of the masses of impacting NEOs. This will enable any correlation between the impact parameters such as flash temperature and duration with the masses of impactors, and will broaden our knowledge about the energy partitioning mechanisms. 

\section*{Acknowledgements}
This work has made use of data from the European Space Agency (ESA) NELIOTA project. Thanks to Marco Delbo' (Observatoire de la C\^ote d'Azur) for his crucial help and important input in the black body and temperature modelling. We would like to thank the anonymous reviewer for his/her comments and suggestions, which led to the significant improvement of the manuscript. CA was supported by the French National Research Agency Fellowship under the project ``Investissements d'Avenir'' UCA$^{JEDI}$ with the reference number ANR-15-IDEX-01 and partially by the CSI (Cr\'edits Scientifiques Incitatifs) grant of the Universit\'e Nice Sophia-Antipolis. CA thanks the European Space Agency for the Research Fellowship in Space Science (2016-2018). Thanks to Danielle Moser and Robert Suggs (NASA Marshall Space Flight Center) for being always available to answer questions. 
\bibliographystyle{mnras}
\bibliography{references.bib} 
\bsp	
\label{lastpage}
\end{document}